\documentstyle[emulateapj,danonecolfloat]{article}
\input psfig.sty

\def\eg{{\it e.g. }}
\def\etal{{\it et al. }}

\begin{document}
\twocolumn[


\articleid{11}{14}

\submitted{August 17, 2005}

\title{On Random Walks with a General Moving Barrier}
\author{Jun Zhang and Lam Hui}
\affil{Department of Physics, Columbia University,
New York, NY 10027\\
{\tt jz203@columbia.edu, lhui@astro.columbia.edu}}
                                          
\begin{abstract}
Random walks with a general, nonlinear barrier have found recent applications ranging
from reionization topology to refinements in the excursion set theory of halos. 
Here, we derive the first-crossing distribution of random walks 
with a moving barrier of an arbitrary shape.
Such a distribution is shown to satisfy an integral equation that can be solved by a simple matrix inversion, 
without the need for Monte Carlo simulations, making this useful for exploring a large parameter space.
We discuss examples in which common analytic approximations fail, a failure which can be
remedied using the method described here. 
\end{abstract}


]

\section{Introduction}
\label{intro}

\begin{figure}[tb]
\centerline{\epsfxsize=9cm\epsffile{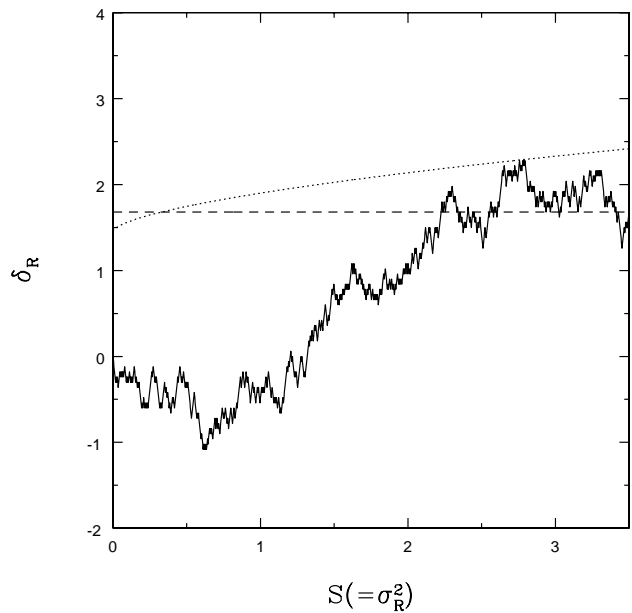}}
\caption{An example of the random walk pattern in the excursion
set theory. $S$ (or $\sigma_R^2$) denotes 
the variance of $\delta_R$ which is the density fluctuation smoothed 
on scale $R$. A large $S$ is equivalent to a small $R$. 
The dashed curve represents a flat barrier, motivated by spherical collapse.
The dotted curve represents a moving barrier, motivated, for instance,
by ellipsoidal collapse.}
\label{randomwalk}
\end{figure}

A random walk is a stochastic process consisting of a sequence of uncorrelated discrete steps. 
It has found applications in diverse areas ever since Einstein's study of the Brownian motion. 
In cosmology, it has been used mainly to model the statistics of halo formation and mergers, 
in a theory sometimes referred to as extended-Press-Schechter, or
excursion set (see \cite{ps74,bcek91,lc93}).
The idea is quite simple. Take a snapshot of the universe at some early time, with
some given initial density field. Pick a point anywhere in the universe. 
Imagine smoothing the density field around this point with a filter, say a top-hat filter,
of a successively smaller radius. Naturally, (in a hierarchical universe such as our own) one expects that the larger
the radius of the filter, the smaller the mean overdensity would be within the filter.
This statement is true on average. For any given realization, there will be fluctuations
as one varies the radius. In other words, a plot of the 
smoothed density versus the size of the filter traces out a random walk, as illustrated in Fig. \ref{randomwalk}. 
The overall trend is for the smoothed density to rise with a smaller filter radius, 
but with significant fluctuations.

The excursion set theory postulates a flat threshold or barrier (usually motivated by spherical collapse):
once the random walk crosses the barrier, a halo is declared to have formed with a size given by the
crossing radius. Here, the filter radius $R$ is often denoted by alternative equivalent quantities: mass $M$ or
variance $S$. The latter is particularly useful for Gaussian random initial conditions -- note that 
$S$ increases as the radius or mass is decreased. 
A natural question to ask is: given Gaussian random initial conditions, what is the probability that
a random walk first crosses the barrier between $S$ and $S + dS$, denoted by
$f(S) dS$? The quantity $f(S)$ is known as
the first-crossing distribution. In excursion set theory, the first-crossing distribution is directly
related to the halo mass function, a key quantity of interest.

Recently, researchers have found it useful to consider random walks with a non-flat barrier -- 
following common practice, we will refer to this as a moving barrier, 
in the sense that the barrier 'moves' with $S$. 
\footnote{The term moving barrier could be confusing to some readers: even in the
case of a flat barrier as in the excursion set theory, the barrier could move in the sense
that it changes with redshift. This is not the sense in which we use the term.}
For instance, Sheth {\it et al.} (2001, \cite{smt01} hereafter) found that an ellipsoidal collapse model suggests
a moving barrier for defining halos, which produces a first-crossing distribution, or halo mass function,
that better matches N-body simulations. 
More generally, random walks can be used to model distributions of objects other than dark matter halos.
For instance, recently, Furlanetto {\it et al.} (2004, \cite{fzh04} hereafter) used the first-crossing distribution of
random walks with a moving barrier to model the HII bubble size distribution in the reionization epoch.

Existing techniques to analytically compute the first-crossing distribution for a flat, or even
linear, barrier cannot be easily generalized to a barrier of an arbitrary shape.
In this paper, we take an alternative approach, in which 
the first-crossing distribution is shown to satisfy a simple integral
equation for any kind of barrier. This equation can be solved in a straightforward manner
by inverting a triangular matrix. Not having to resort to Monte Carlo simulations is useful especially
when exploring a large parameter space.
We also show that the well-known analytic flat/linear solutions are reproduced in this approach.
This is done in \S\ref{main}.
In \S\ref{application}, we briefly describe a few astrophysical applications, and show under what circumstances
(non-Monte-Carlo) analytic approximations that are currently in use might fail, a failure for which our
approach provides a remedy.

\section{Random walks with a moving barrier}
\label{main}
In this section, we calculate the first-crossing distribution of random walks with a moving barrier. The distribution satisfies an integral equation which can be solved analytically only when the barrier is linear. In general, the integral equation can be solved by inverting a triangular matrix. The resulting solution is shown to agree well with Monte Carlo simulations.
 
\subsection{The First-Crossing Distribution}
\label{rw}
In a 1-D random walk model, the probability distribution of the displacement $\delta$ (which is equivalent to
the smoothed density in the excursion set theory) is Gaussian. We use $S$ to stand for the variance of the displacement. 
The moving barrier is denoted by $B(S)$. The probability that the random walk first crosses the barrier at 
between $S$ and $S+dS$ is represented by $f(S)dS$. Furthermore, we define $P(\delta,S)d\delta$ as the probability that the random walk crosses between $\delta$ and $\delta+d\delta$ at $S$ without ever crossing the barrier before $S$
(i.e. less than $S$). 
Since the random walk either crosses the barrier before $S$ or passes $(\delta,S)$ with $\delta<B(S)$, we have:
\begin{equation}
\label{sum}
1=\int_0^Sf(S')dS'+\int_{-\infty}^{B(S)}P(\delta,S)d\delta
\end{equation}

Ignoring the barrier, $P(\delta,S)$ should be equal to $P_0(\delta,S)$ which is the normal Gaussian distribution with variance $S$ and defined as:
\begin{equation}
\label{Gaussian}
P_0(\delta,S)=\frac{1}{\sqrt{2\pi S}}\exp (-\frac{\delta^2}{2S})
\end{equation}
When there is a barrier, to get $P(\delta,S)$, one should subtract from $P_0(\delta,S)$ the probability that the random walk crosses the barrier at somewhere before $S$ and subsequently passes through $(\delta,S)$. Therefore:
\begin{equation}
\label{definitionOfP}
P(\delta,S)=P_0(\delta,S)-\int_0^SdS'f(S')P_0(\delta-B(S'),S-S')
\end{equation} 

We are now ready to solve for both $f(S)$ and $P(\delta,S)$ as there are two integral equations. Taking
the derivative of eq.~[\ref{sum}] with respect to $S$, we get:
\begin{equation}
\label{No1}
f(S)=-P(B(S),S)\frac{dB}{dS}-\int_{-\infty}^{B(S)}\frac{\partial P(\delta,S)}{\partial S}d\delta
\end{equation}
Using eq.~[\ref{definitionOfP}] in eq.~[\ref{No1}], we obtain an integral equation for the first-crossing distribution $f(S)$,
the key equation of our paper:
\begin{equation}
\label{f}
f(S)=g_1(S)+\int_0^SdS'f(S')g_2(S,S')
\end{equation}
where 
\begin{equation}
\label{g1}
g_1(S)=(\frac{B(S)}{S}-2\frac{dB}{dS})P_0(B(S),S)
\end{equation}
and
\begin{equation}
\label{g2}
g_2(S,S')=[2\frac{dB}{dS}-\frac{B(S)-B(S')}{S-S'}]P_0(B(S)-B(S'),S-S')
\end{equation}

To reach eq.~[\ref{f}], one needs to use the following two relations:
\begin{equation}
\label{trick}
[\int_{-\infty}^{B(S)}d\delta P_0(\delta-B(S'),S-S')]\vert_{S -> S'}=\frac{1}{2}
\end{equation}
and
\begin{equation}
\label{tric}
\int_{-\infty}^{B(S)}\frac{\partial P_0(\delta,S)}{\partial S}d\delta = -\frac{B(S)}{2S}P_0(B(S),S)
\end{equation}
Eq.~[\ref{trick}] can be proven using Taylor expansion of $B(S)-B(S')$ around $S=S'$(assuming B(S) is differentiable). 
Eq.~[\ref{tric}] can be derived by noticing that $P_0(\delta,S)$ satisfies the diffusion equation:
\begin{equation}
\label{diffusion}
\frac{\partial P_0}{\partial S}=\frac{1}{2}\frac{\partial^2 P_0}{\partial \delta^2}
\end{equation}

Eq.~[\ref{f}] takes a form sometimes known as the 
Volterra integral equation of the second kind. 
In general, it has a unique solution:
\begin{equation}
\label{solution}
f(S)=g_1(S)+\sum_{n=1}^{\infty}\int_0^SdS'g_1(S')G_n(S,S')
\end{equation}
where
\begin{eqnarray}
\label{defineG}
&&G_1(S,S')=g_2(S,S')\\ \nonumber
&&G_{n+1}(S,S')=\int_{S'}^Sg_2(S,u)G_n(u,S')du
\end{eqnarray}

\subsection{Linear Barrier}
If the barrier $B(S)$ is a linear function of $S$, $g_2(S,S')$ is a function only of $S-S'$. The solution of eq.~[\ref{f}] can be 
written down in closed form using Laplace transformation. 
The Laplace transformation is defined as:
\begin{eqnarray}
\label{Laplace}
&&\tilde{f}(t)=\int_0^{\infty}\exp (-St)f(S)dS \\ \nonumber
&&\tilde{g}_1(t)=\int_0^{\infty}\exp (-St)g_1(S)dS \\ \nonumber
&&\tilde{g}_2(t)=\int_0^{\infty}\exp [-(S-S')t]g_2(S-S')d(S-S')
\end{eqnarray}
Eq.~[\ref{f}] implies that
\begin{equation}
\tilde{f}(t)=\tilde{g}_1(t)+\tilde{f}(t)\tilde{g}_2(t)
\end{equation}
Assuming the barrier has the form:
\begin{equation}
B(S)=a+bS
\end{equation}
We have 
\begin{eqnarray}
&&\tilde{g}_1(t)=(1-\frac{b}{\sqrt{b^2+2t}})\exp (-ab-a\sqrt{b^2+2t}) \\ \nonumber
&&\tilde{g}_2(t)=\frac{b}{\sqrt{b^2+2t}}
\end{eqnarray}
Therefore
\begin{equation}
\label{tildef}
\tilde{f}(t)=\exp (-ab-a\sqrt{b^2+2t})
\end{equation}
Inverting eq.~[\ref{tildef}] gives the well-known Inverse Gaussian distribution
(\eg Sheth 1998, Sheth \& Tormen 2002):
\begin{equation}
\label{inverseG}
f(S)=\frac{B(0)}{S\sqrt{2\pi S}}\exp (-\frac{B^2(S)}{2S})
\end{equation}
This confirms that the integral equation [\ref{f}] does yield
the correct solution in the special case of a linear barrier.

\subsection{General Cases}
For a general moving barrier, eq.~[\ref{f}] can be solved numerically 
on a mesh with equal spacing (see Press {\it et al.} 1992 
for more details):
\begin{equation}
\label{spacing}
S_i=i\times\Delta S, i=0,1,...,N, \Delta S=\frac{S}{N}
\end{equation}
The integral equation is effectively a set of linear algebraic equations:
\begin{equation}
\label{lineareq}
f(S_i)=g_1(S_i)+\frac{\Delta S}{2}\sum_{j=1}^{j=i}g_2(S_i,S_j-\frac{\Delta S}{2})[f(S_j)+f(S_{j-1})]
\end{equation}
If we treat $f(S_i)$ as a vector, we have:
\begin{equation}
\label{vector}
\mathbf{F}=\mathbf{G}+\mathbf{M}\mathbf{F}
\end{equation}
where $F_i$ $= f(S_i)$ and $G_i$ $= g_1(S_i)$, both of which are $N+1$ dimensional vectors. $\mathbf{M}$ is an $(N+1)\times(N+1)$ matrix and defined as:
\[ M_{ij}= \left \{ \begin{array}{ll}
0 & \mbox{if $j>i$};\\
\Delta_{i,i} & \mbox{if $j=i \neq 0$};\\
\Delta_{i,j+1}+\Delta_{i,j} & \mbox{if $0<j<i$};\\
\Delta_{i,1} & \mbox{if $j=0$ \& $i \neq 0$};\\
0 & \mbox{if $i=j=0$}.\end{array} \right. \]
where $\Delta_{i,j}=\frac{\Delta S}{2}g_2(S_i,S_j-\frac{\Delta S}{2})$.
We immediately have $\mathbf{F}=(\mathbf{I}-\mathbf{M})^{-1}\mathbf{G}$ as the solution of the integral equation. $\mathbf{I}$ is the identity matrix. Since $\mathbf{M}$ is a triangular matrix, the equation can be solved iteratively in a straightforward way:
\begin{eqnarray}
\label{solveonmesh}
f(S_0)&=&g_1(S_0)=0 \\ \nonumber
f(S_1)&=&g_1(S_1)[1-\Delta_{1,1}]^{-1}\\ \nonumber
f(S_i)\vert_{i>1}&=&[1-\Delta_{i,i}]^{-1}\\ \nonumber
&\times&[g_1(S_i)+\sum_{j=1}^{i-1}f(S_j)(\Delta_{i,j}+\Delta_{i,j+1})]
\end{eqnarray}

There is one complication here: since $g_2(S,S')$ approaches infinity when $S$ approaches $S'$, one needs to define $g_2(S,S-\frac{\Delta S}{2})$ carefully when $\Delta S$ is small. According to eq.~[\ref{g2}], we know:
\begin{equation}
\label{g2limit}
g_2(S,S')\vert_{S -> S'} \sim \frac{1}{\sqrt{S-S'}}
\end{equation}
We re-define $g_2(S,S-\frac{\Delta S}{2})$ in eq.~[\ref{solveonmesh}] as:
\begin{equation}
\label{g2deltalimit}
g_2(S,S-\frac{\Delta S}{2})=\frac{1}{\Delta S}\int_{S-\Delta S}^S g_2(S,S')dS' 
\end{equation}

\begin{figure}[tb]
\centerline{\epsfxsize=9cm\epsffile{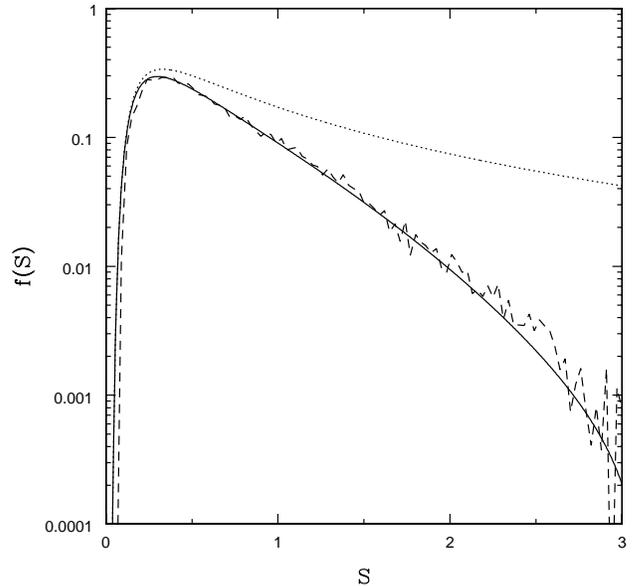}}
\caption{The first-crossing distribution for a barrier of the form: $B(S)=1+0.3S+0.3S^2$. The dashed curve is from a
Monte Carlo simulation; the solid curve is the solution of eq.~[\ref{f}]. For comparison, we also show the dotted curve which is for a linear barrier of the form: $B(S)=1+0.3S$. The discrepancy between the simulation and our exact solution 
at very small $S$ is due to the small number of random walks that first cross at such an $S$ in the simulation.
}
\label{compare}
\end{figure}

To compare our calculation with a Monte Carlo simulation, the barrier is chosen to be $B(S)=1+0.3S+0.3S^2$. In Fig. \ref{compare}, the solid curve is the numerical solution of eq.~[\ref{f}] using eq.~[\ref{solveonmesh}].
For comparison, the dotted curve is the first-crossing distribution of a linear barrier of the form $B(S)=1+0.3S$
-- a nonlinear barrier is sometimes approximated as a linear one in the literature.
The dashed curve is from a Monte Carlo 
simulation. Our integral equation approach and the Monte Carlo simulation yield consistent results.
The former is naturally less noisy.

\section{Applications In Cosmology}
\label{application}
In this section, we discuss the cosmological applications of random walks with a moving barrier through two examples: the halo mass function in the ellipsoidal collapse model and the HII bubble size distribution during reionization. We will see that 
an analytic fitting formula for the first-crossing distribution proposed by Sheth \& Tormen (2002, \cite{st02} hereafter) is a good approximation when the barrier is mildly nonlinear. However, the fitting formula can significantly differ from the exact solution for a more general moving barrier.     

\subsection{The Ellipsoidal Collapse Model}
As explained in the introduction,
in the usual formulation of the excursion set theory, the halo mass function is given by the first-crossing distribution of random walks with a flat barrier whose height is determined by the spherical collapse model: 
\begin{equation}
\label{massfunction}
n(m)dm=\frac{\overline{\rho}}{m}f(S)dS
\end{equation}
where $n(m)dm$ is the number density of dark matter halos with mass between $m$ and $m+dm$; $f(S)$ is the first-crossing distribution; $S$ is the variance of the mass overdensity at scale $m$; $\overline{\rho}$ is the mean mass density. This Press-Schechter mass function agrees with numerical simulations reasonably well, but a significant discrepancy is found for small halos (see \cite{smt01} for details). 

\cite{smt01} points out that such a discrepancy at small scales may result from the fact that the spherical collapse model is over-simplified. Since in general, peaks in a Gaussian random field are ellipsoidal (\cite{d70,bbks86}), they argue that the formation of halos should be better described by an ellipsoidal collapse model, in which a moving barrier should be used instead of a flat barrier. The barrier 
adopted by \cite{smt01} is:
\begin{equation}
\label{STbarrier}
B(S)=\sqrt{a}\delta_{sc}[1+\beta (a\nu)^{-\alpha}]
\end{equation}
where $\nu\equiv\delta_{sc}^2/S$; $\delta_{sc}$ is the critical overdensity 
as in the spherical collapse model; the constant parameters are: $\beta\approx 0.485$, $a\approx 0.75$ and $\alpha\approx 0.615$. Note that $\delta_{sc}$ is a function of redshift $z$.
It is found in numerical simulations that the ellipsoidal collapse model provides
a significant improvement over the spherical collapse model.

To find out the first-crossing distribution of a random walk model with a moving barrier
of the form eq.~[\ref{STbarrier}], \cite{smt01} suggests a fitting formula which is based on Monte Carlo simulations. \cite{st02} proposes a more general fitting formula for moving barriers:
\begin{equation}
\label{st02}
f(S)dS=\vert T(S)\vert \exp [-\frac{B(S)^2}{2S}]\frac{dS/S}{\sqrt{2\pi S}}
\end{equation}
where
\begin{equation}
\label{TS}
T(S)=\sum_{n=0}^5 \frac{(-S)^n}{n!}\frac{\partial^nB(S)}{\partial S^n}
\end{equation}
and $B(S)$ is the barrier. It is obvious that eq.~[\ref{st02}] gives the exact first-crossing distribution for a linear barrier. 

\begin{figure}[tb]
\centerline{\epsfxsize=9cm\epsffile{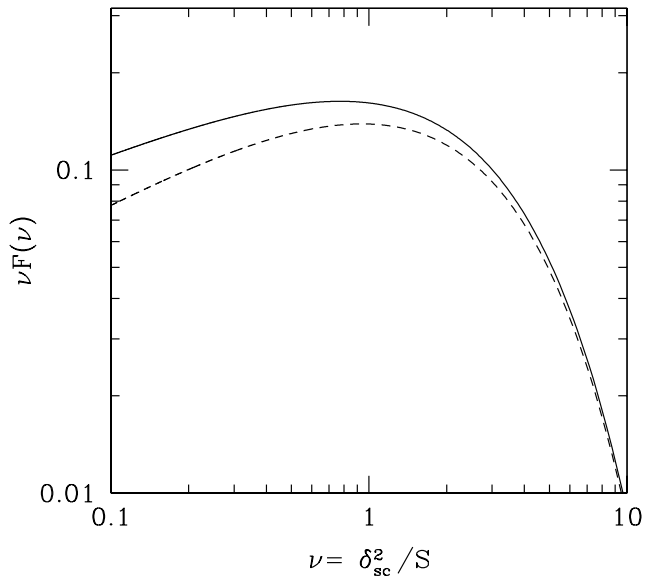}}
\caption{The first-crossing distribution for a barrier of the form given by eq.~[\ref{STbarrier}], which is
motivated
by ellipsoidal collapse. The solid curve is the exact solution from solving eq.~[\ref{f}]; the dashed curve is the fitting formula of ST02, eq.[\ref{st02}].}
\label{ST}
\end{figure}

We compare eq.~[\ref{st02}] with the exact solution of eq.~[\ref{f}] in Fig. \ref{ST}. Instead of using $f(S)$ to represent the first-crossing distribution, it turns out to be convenient to
change variable to $\nu$ and define:
\begin{equation}
\label{nu}
F(\nu)d\nu=f(S)dS
\end{equation}
The fact that $\delta_{sc}$ is a function of redshift $z$ implies that $f(S)$ is implicitly a function of $z$
as well. The function $F$, on the other hand, has the virtue of being independent of $z$, as can
be seen from either eq.~[\ref{f}] or eq.~[\ref{st02}].
Fig. \ref{ST} shows that
the fitting formula does provide a good approximation to the first-crossing distribution
in this case. The relative error starts to grow larger than ten percent when $\nu < 1 $, which corresponds to halo mass less than 
about $10^{13}M_{\odot}$ at $z=0$.

\subsection{The HII Bubble Size Distribution During Reionization}

The evolution of HII bubbles around the epoch of reionization has been studied using both semi-analytic models (\eg \cite{b02,hh03,lbh05}) and numerical simulations (\eg \cite{g00,sahs03,csw03,syahs04}). More recently, \cite{fzh04} proposes a model for such HII bubbles based on
the excursion set theory.
It assumes that the amount of neutral gas being ionized in the bubble is directly proportional to the amount of mass in the collapsed objects with a virial temperature larger than $10^4K$. In other words, the ionized fraction is proportional to the collapsed fraction. To form an HII bubble, one requires the ionized fraction to be equal to one, which corresponds to a critical collapsed fraction. With the excursion set model, one can relate the collapsed fraction with the mass overdensity of the bubble. Thus the critical collapsed fraction actually refers to a critical overdensity for the formation of HII bubbles. The random walk model used in the calculation of the halo mass function can be completely carried over to determine the HII bubble size distribution. According to \cite{fzh04}, the barrier can be written as:
\begin{equation}
B(S)=\delta_{sc} - A[\sigma^2(m_{min})-S]^{\frac{1}{2}}
\label{BFZH}
\end{equation}
where $S$ is $\sigma^2(m)$, the variance of the mass overdensity of bubbles on scale mass $m$; $A$ is a constant determined by the radiation efficiency; $m_{min}$ is the mass of dark matter halos with a virial temperature of $10^4K$, which is about $10^8M_{\odot}$ at redshift 20; $\delta_{sc}$ is the usual critical overdensity in the spherical collapse model. Note that, as before, $B(S)$ is an implicit function of the redshift $z$, through $\delta_{sc}$ and
$m_{min}$.
The HII bubble size distribution is:
\begin{equation}
\label{bubblesize}
n_{HII}(m)dm=\frac{\overline{\rho}}{m}f(S)dS
\end{equation}
where $n_{HII}(m)$ is the number density of bubbles with mass $m$; $\overline{\rho}$ is the mean mass density; $f(S)$ is the first-crossing distribution for the barrier in eq.~[\ref{BFZH}].

\begin{figure}[tb]
\centerline{\epsfxsize=9cm\epsffile{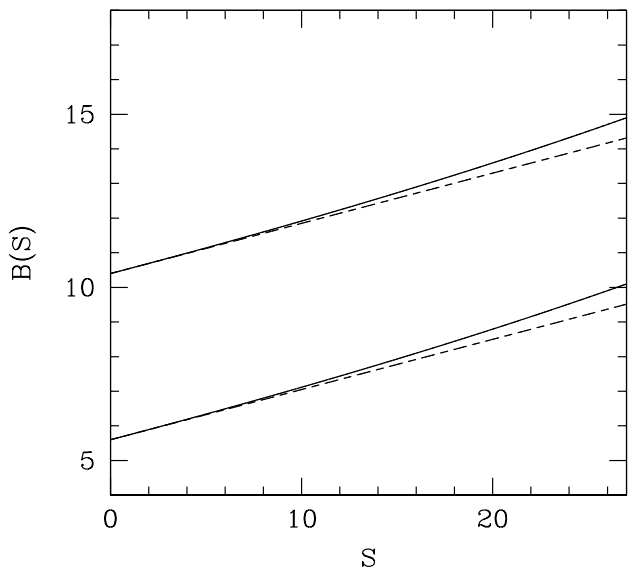}}
\caption{The shapes of the barriers assumed in \cite{fzh04} for different redshifts. The upper set is at $z=20$, and 
the lower set is at $z=16$. The solid lines are the true barrier shapes, and the dashed lines are the linear approximations.  
}
\label{barrierFZH}
\end{figure}

Due to the lack of a non-Monte-Carlo method of treating random walks with a moving barrier, \cite{fzh04} uses a linear fit of the true barrier instead:
\begin{equation}
\label{FZHlinear}
B(S)=B_0+B_1S
\end{equation}
where $B_0=\delta_{sc}-A\sigma(m_{min})$ and $B_1=\frac{A}{2\sigma(m_{min})}$. Fig. \ref{barrierFZH} shows the shapes of both the true barrier and its linear fit at redshift $z=20,16$ respectively. 

\begin{figure}[tb]
\centerline{\epsfxsize=9cm\epsffile{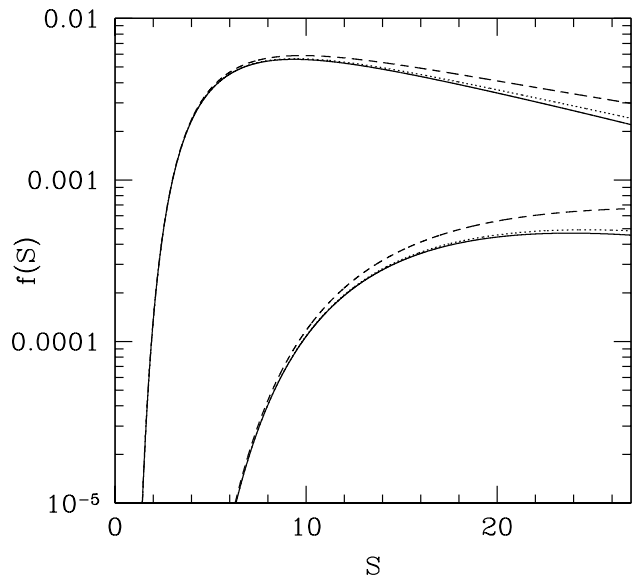}}
\caption{The first-crossing distribution for the barriers shown in Fig. \ref{barrierFZH}. The lower set is at $z=20$; the upper set is at $z=16$. The dashed curves are for the linear barriers. Both the solid and the dotted curves are for the true barriers. The solid lines are the exact solutions; the dotted lines are from the fitting formula given by eq.~[\ref{st02}].}
\label{FZH}
\end{figure}

We calculate the first-crossing distribution in three ways: one is to use eq.~[\ref{f}], which gives the exact solution; the second way is to use eq.~[\ref{st02}], the fitting formula given by \cite{st02}; we also include the first-crossing distribution for the linear barrier defined in eq.~[\ref{FZHlinear}]. The results are presented in Fig. \ref{FZH}. One can see all three methods yield similar results. The fitting formula of \cite{st02} in fact
provides a surprisingly good approximation to the exact solution, with a maximum error of less than 10 percent
for the relevant range of $S$. The linear barrier approximation is less accurate, especially at a large $S$.

\subsection{Cautionary Remarks on the Fitting Formula}
\begin{figure}[tb]
\centerline{\epsfxsize=9cm\epsffile{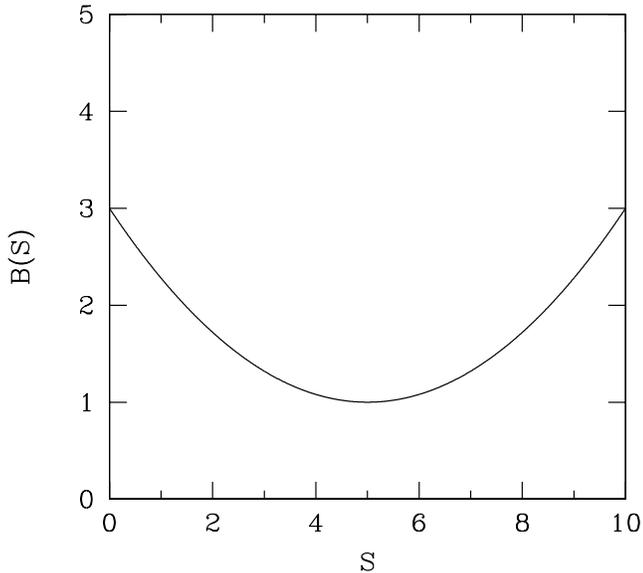}}
\caption{A hypothetical example of a complex barrier: $B(S)=1+\frac{2}{25}(S-5)^2$.}
\label{complexbarrier}
\end{figure}

\begin{figure}[tb]
\centerline{\epsfxsize=9cm\epsffile{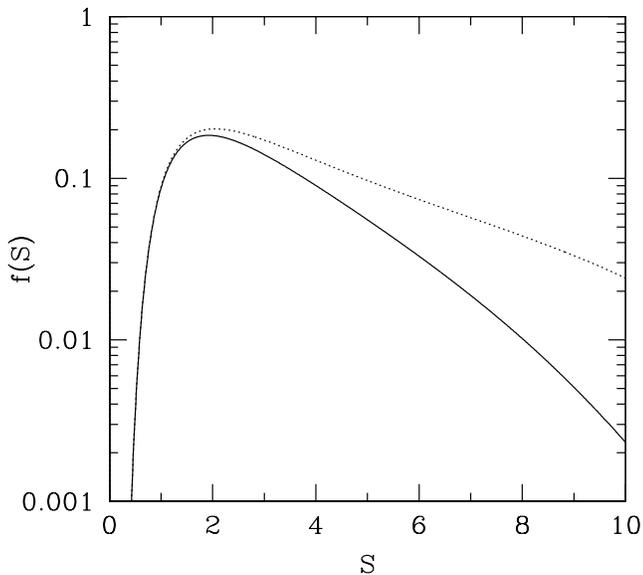}}
\caption{The first-crossing distribution of random walks with a moving barrier shown in Fig. \ref{complexbarrier}. The solid curve is the exact solution; the dotted curve is the fitting formula of \cite{st02}.}
\label{counterexample}
\end{figure}

The discussion above suggests that the fitting formula by \cite{st02} works well when the barrier is not
too complicated (i.e. only mildly nonlinear). However, when the barrier has a complex shape, the fitting formula may fail. We show a hypothetical example in Fig. \ref{counterexample}. The barrier is chosen to be $B(S)=1+\frac{2}{25}(S-5)^2$, which is shown in Fig. \ref{complexbarrier}. In this example, the first-crossing distribution given by the fitting formula differs significantly from the exact solution.

Non-monotonic barriers such as the one above likely
have useful applications in cosmological problems. 
Consider again for instance the problem of the HII bubble size distribution during reionization. 
In \cite{fzh04}, the barrier has a simple monotonic form in part because recombination is ignored.
Recently, Furlanetto \& Oh (2005) point out that recombination disfavors very large HII bubbles, with
an upper limit set roughly by the mean free path of the ionizing photons. 
Since the mean free path depends on the average overdensity of a bubble, and since the 
rms density fluctuation is a function of scale (i.e. S),
it is conceivable that recombination would modify the barrier shape 
from those shown in Fig. \ref{barrierFZH} by
causing an upturn at small $S$ (suppressing large bubbles), perhaps similar to the one in our hypothetical
example (Fig. \ref{complexbarrier}).
Moreover, one should keep in mind that ionizing sources associated with different mass scales (and therefore
different $S$'s) might have quite different radiation efficiencies, leading also to the possibility of
a complex barrier shape for HII bubble formation. Going into details of these models would be
beyond the scope of this paper. It suffices to point out that there are situations in which
commonly used analytic approximations, whether it be a simple linear approximation or the formula proposed by \cite{st02},
are inadequate. Our approach via solving eq.~[\ref{f}] by matrix inversion might provide a useful alternative, avoiding
the need for Monte Carlo simulations which can be time consuming when exploring a large parameter space.

\section{Summary}
\label{summary}
To summarize, we show
that the first-crossing distribution of random walks with a general moving barrier satisfies 
eq.~[\ref{f}], the key equation in this paper. 
This integral equation can be solved by inverting a triangular matrix. A simple iterative scheme is presented in
eq.~[\ref{solveonmesh}]. We verify the technique by comparing with Monte Carlo simulations.
We also show that the integral equation nicely yields the well-known solution for the linear barrier case.
Some cosmological applications are briefly discussed, and we caution that existing analytic approximations
might fail in cases with a complex barrier (such as a non-monotonic one), where our approach might prove useful.

Research for this work is supported in part by the DOE DE-FG02-92-ER40699 and the NSF AST-0098437.

\vskip 1cm

\end{document}